\documentstyle[twoside,fleqn,espcrc2,epsfig]{article}

\newcommand{\AmS}{{\protect\the\textfont2
  A\kern-.1667em\lower.5ex\hbox{M}\kern-.125emS}}

\title { Effective Mass Generation of Off-diagonal Gluons 
and Abelian Dominance in the Maximally Abelian Gauge in QCD}
\author{K.~Amemiya and H.~Suganuma
\address{Research Center for Nuclear Physics (RCNP), Osaka University, 
Mihogaoka 10-1, Ibaraki, Osaka 567-0047, Japan}} 

\def\be{\begin{equation}}
\def\ee{\end{equation}}
\def\bea{\begin{eqnarray}}
\def\eea{\end{eqnarray}}
\def\b{\begin{eqnarray*}}
\def\e{\end{eqnarray*}}
\def \no{\nonumber}
\def \( {\left(}
\def \) {\right)}

\def\[{\left[}
\def\]{\right]}
\def\lsim {~^{<~}_{\sim~}}
\def\gsim {~^{>~}_{\sim~}}
\def\d{\partial}
\def\f{\frac}

\def\Gmumu{G_{\mu\mu}}
\def\Amu{A_{\mu}}

\def\Meff{M_{\rm off}}
\def \hDdmu {\hat{D}_{\mu}}%

\begin{document}
\begin{abstract}
{\normalsize
We study the properties of gluons in QCD
in the maximally abelian (MA) gauge.
In the MA gauge,
the off-diagonal gluon behaves
as the massive vector boson with the mass $\Meff \simeq 1.2 {\rm~ GeV}$,
and therefore the off-diagonal gluon cannot carry the long-range interaction 
for $r \gg \Meff^{-1} \simeq 0.2$ fm.
The essence of the infrared abelian dominance in the MA gauge
is physically explained
with the generation of the off-diagonal gluon mass $\Meff \simeq 1.2 {\rm~ GeV}$
induced by the MA gauge fixing,
and the off-diagonal gluon mass generation would predict
general infrared abelian dominance  in QCD in the MA gauge.
We report also the off-diagonal gluon propagator at finite temperature.}
\vspace{-.2cm}
\end{abstract}

\maketitle
\section{Introduction}
The quark-confinement mechanism is one of the most
important subjects in the nonperturbative QCD.
In the dual-superconductor picture \cite{YN},
the quark-confinement mechanism can be physically interpreted
with the dual Meissner effect.  
In this context,
the MA gauge has been paid much attention to 
construct the dual-superconductor picture from QCD \cite{3,EI,5}.
In the MA gauge, diagonal gluon components 
behave as neutral gauge fields like photons, 
and off-diagonal gluon components behave as charged matter fields 
on the residual abelian gauge symmetry. 
In the MA gauge, 
the diagonal part of the gluon field has been demonstrated to 
play a dominant role to the nonperturbative quantities 
like confinement \cite{2} and chiral symmetry breaking  \cite{OM}. 
On the other hand, the off-diagonal part of the gluon field  
does not contribute to the long-range phenomena. 
This, abelian dominance \cite{EI}, 
is one of the key concepts to link the dual-superconductor picture from QCD.

\section{Mass-Generation Hypothesis on Off-diagonal Gluons in the MA gauge}
Abelian dominance for the infrared region, which we call 
``infrared abelian dominance'', 
is usually discussed on the role of
the diagonal component of the gluon field.
However, in terms of the off-diagonal gluon,
infrared abelian dominance can be expressed that off-diagonal gluon components
are inactive at the infrared scale of QCD
and can be neglected for the argument of the nonperturbative QCD. 
As a possible physical interpretation for infrared abelian dominance
as infrared inactivity of off-diagonal gluon,
we conjecture that the effective mass of the off-diagonal gluon
$\Amu^\pm \equiv \f{1}{\sqrt 2}( \Amu^1 \pm i \Amu^2)$
is induced in the MA gauge.
If $\Amu^\pm$ acquires a large effective mass $\Meff$,
the off-diagonal gluon propagation is limited within the short range
as $r \lsim \Meff^{-1}$.
This mass-generation hypothesis on off-diagonal gluon
in the MA gauge is formally expressed in QCD as follows \cite{2,6,7}.
The SU(2) QCD partition functional in the MA gauge is expressed as
\bea
Z_{\rm QCD}^{\rm MA}  
\!\!\!\!\!&=& \!\!\!\!\!\!\!
\int \!\!DA_\mu e^{iS_{\rm QCD}[A_\mu ]}
\delta (\Phi _{\rm MA}^\pm [A_\mu ])\Delta _{\rm PF}[A_\mu ],
\eea
where $\Delta_{\rm FP}[\Amu]$ denotes
the Faddeev-Popov determinant. 
Here, $\Phi_{\rm MA}^{\pm}[\Amu]$
denotes the off-diagonal component of
$
\Phi_{\rm MA}[\Amu]
        \equiv
        [\hDdmu,[\hat D^\mu,\tau^3 ]] 
$ which is diagonalized in the MA gauge 
\cite{2,7,ichiead}, 
and therefore the MA gauge fixing is provided by
$\delta(\Phi_{\rm MA}^{\pm}[\Amu])$.
The mass-generation hypothesis of off-diagonal
gluon $\Amu^{\pm}$ is expressed as
\vspace{-.1cm}
\bea
Z^{\rm MA}_{\rm QCD}
\!\!\!\!&=& \!\!\!\!\!
\!\!\int\!\!\!{\cal D}\Amu^3
                        e^{ iS_{\rm Abel}[\Amu^3] }
                \!\!\int\!\!\!{\cal D}\Amu^{\pm}
                        e^{ iS_{\rm off}^{M}[\Amu^\pm]}
                {\cal F} [\Amu].
                                                        \label{eqn:4simc01}
\eea
Here, $S_{\rm off}^{M}[\Amu^\pm]$ denotes
the U(1)$_3$-invariant action of the off-diagonal gluon
with the effective mass $\Meff$ as
\bea
S_{\rm off}^{M}[\Amu^\pm]\!\!\!\!\!\!& \equiv&  \!\!\!\!\!\!
        \int d^4 x
        \biggl\{ -\f{1}{2}
           \( D^{\rm Abel}_\mu A_\nu^+ - D^{\rm Abel}_\nu A_\mu^+ \) \no\\
\times  
( \!\!\!\!\!\!&D_{\rm Abel}^{\mu*}&\!\!\!\!\!\!A^\nu_- - D_{\rm Abel}^{\nu*} A^\mu_- )
           \!+\! \Meff^2 \Amu^+A^\mu_-
         \biggr\} 
                                                        \label{eqn:4simc02}
\eea
with the U(1)$_3$ covariant derivative  
$D^{\rm Abel}_\mu \equiv \d_\mu +ieA^3_\mu$.
Here, $S_{\rm Abel}[\Amu^3]$ is the effective action
of the diagonal gluon component,
and
${\cal F}[\Amu]$ is a U(1)$_3$-invariant smooth functional
in comparison with
$\exp \left\{ iS_{\rm off}^{M}[\Amu^\pm] \right\}$ at least
in the infrared region.
Since the effective mass $\Meff$ is closely related to 
the off-diagonal gluon propagation,
we study the gluon propagator in the MA gauge
in terms of the interaction range
using the SU(2) lattice QCD simulations \cite{2,6,7}.
%

%
\begin{figure}[h]
\vspace{-0.6cm}
\begin{center}
\epsfig{figure=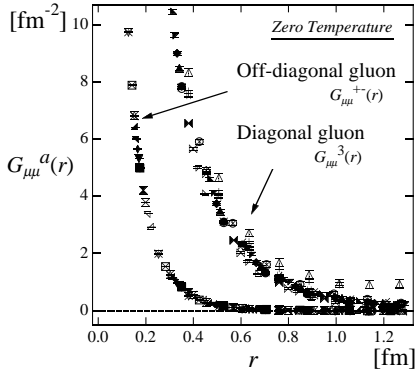,height=5.0cm}
\vspace{-1.1cm}
\caption
{ The scalar-type gluon propagator $\Gmumu^a(r)$ 
as the function of the 4-dimensional
distance $r$ in the MA gauge 
with $2.2 \le \beta \le 2.4$ and various lattice size
($12^3 \times 24$, $16^4$, $20^4$).
}\label{cap:gpT0}
\end{center}
\vspace{-1.3cm}
\end{figure}

\begin{figure}[htb]
\vspace{-.2cm}
\begin{center}
\epsfig{figure=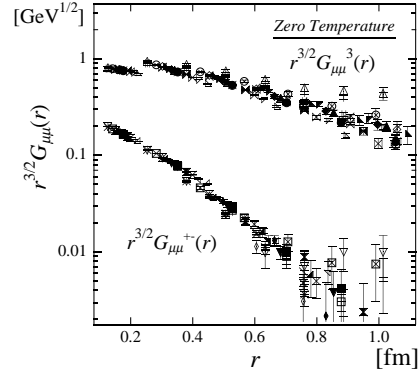,height=5.0cm}
\vspace{-1.1cm}
\caption{
 The logarithmic plot for
the scalar correlation $r^{3/2} G_{\mu \mu}^a(r)$.
The off-diagonal gluon propagator behaves as
the Yukawa-type function,
$G_{\mu \mu }^{+-} \sim {\exp(-\Meff r) \over r^{3/2}}$.
}\label{cap:r2i3G}
\vspace{-0.9cm}
\end{center}
\end{figure}

\section{Effective Mass of the Off-diagonal Gluon in the MA Gauge}
In the Euclidean QCD, the MA gauge is 
defined by minimizing 
$R_{\rm off} [A_\mu] \equiv 
{e^2} \int d^4x A_\mu^+(x)A^\mu_-(x).
$ 
In the MA gauge, the off-diagonal gluon components 
are forced to be as small as possible by the SU(2) 
gauge transformation \cite{2,ichiead}.  
In the MA gauge, the SU(2) gauge symmetry 
is reduced into the ${\rm U(1)_3}$ gauge symmetry.
As for the residual U(1)$_3$ gauge symmetry, 
we impose the U(1)$_3$ Landau gauge fixing 
to extract most continuous gauge configuration 
under the MA gauge constraint 
and to compare with the continuum theory \cite{2,6,7}.

To investigate the off-diagonal gluon mass $\Meff $, 
we study the Euclidean gluon propagator 
$G_{\mu \nu }^{ab} (x-y) \equiv \langle A_\mu ^a(x)A_\nu ^b(y)\rangle$ 
in the MA gauge with the U(1)$_3$ Landau gauge fixing \cite{2,6,7}.
The continuum gluon field $A_\mu^a(x)$ is extracted 
 from the link variable as 
$U_\mu(s)={\rm exp}(iaeA_\mu^a(s) \frac{\tau^a}{2})$.
Here, the scalar-type gluon propagator 
$G_{\mu\mu}^a(r)\equiv \sum^4_{\mu=1}\langle 
A_\mu^{~a}(x)A_\mu^{~a}(y)\rangle$ 
is useful to observe the interaction range of the gluon,
because it depends only on the four-dimensional Euclidean 
distance $r \equiv \sqrt{(x_\mu- y_\mu)^2}$.

We show in Fig.\ref{cap:gpT0} $G_{\mu \mu}^3(r)$ and 
$
G_{\mu\mu}^{+-}(r) \equiv 
\sum^4_{\mu=1}\langle A_\mu^{+}(x)A_\mu^{-}(y)\rangle
={1 \over 2} \{G_{\mu\mu}^1(r)+G_{\mu\mu}^2(r)\}
$
in the MA gauge with the U(1)$_3$ Landau gauge fixing.
In the MA gauge, $\Gmumu^{\rm 3}(r)$ and
$\Gmumu^{\rm +-}(r)$ manifestly differ.
The diagonal-gluon propagator $\Gmumu^{\rm 3}(r)$
takes a large value even at the long distance.
In fact, the diagonal gluon $\Amu^3$ in the MA gauge
propagates over the long distance.
On the other hand, the off-diagonal gluon propagator
$\Gmumu^{\rm +-}(r)$ rapidly decreases and is negligible
for $r \gsim 0.4$ fm in comparison with $\Gmumu^{\rm 3}(r)$.
Then, the off-diagonal gluon $\Amu^\pm$ seems to propagate
only within the short range as $r \lsim 0.4$ fm.
Thus, infrared abelian dominance for the gluon propagator
is found in the MA gauge.
Since the massive vector-boson propagator with the mass $M$ 
takes a Yukawa-type asymptotic form as $G_{\mu\mu}(r) \sim 
{M^{1/2} \over r^{3/2}}\exp(-Mr)$ for $Mr > 1$, 
the effective mass $\Meff $ of the off-diagonal gluon 
$A_\mu^{\pm}(x)$ can be evaluated from the slope of the logarithmic 
plot of $r^{3/2} G_{\mu\mu}^{+-}(r)\sim \exp(-\Meff r)$ 
as shown in Fig.\ref{cap:r2i3G}. 
The off-diagonal gluon $A_\mu^{\pm}(x)$ 
behaves as the massive vector boson with $\Meff  \simeq 1.2~{\rm GeV}$ 
in the MA gauge for $r \gsim 0.2$ fm \cite{2,6,7}. 

\section{Gluon Propagator in the MA Gauge at Finite Temperature}
\begin{figure}[tb]
\vspace{-.2cm}
\begin{center}
\epsfig{figure=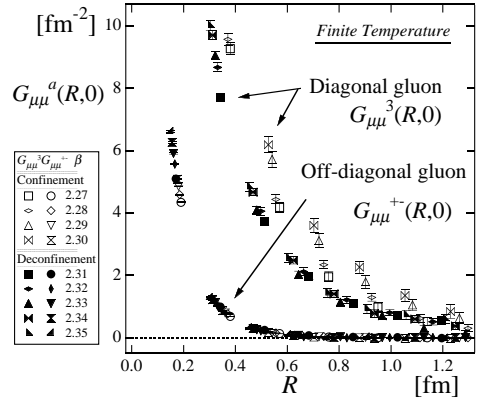,height=5.3cm}
\vspace{-1.1cm}
\caption{
The scalar-type gluon spatial correlation $\Gmumu^{a}(R,0)$ at finite temperatures 
as the function of the 3-dimensional distance $R$ in the MA gauge 
with $2.27 \le \beta \le 2.35$ on $12^2 \times 4 \times 24$ lattice.
}
\label{caption:Gft}
\end{center}
\vspace{-0.9cm}
\end{figure}

We study also the scalar-type gluon propagator in the MA gauge
at finite temperature \cite{finitethesis}.
We measure the spatial correlation of 
$
\Gmumu^{a}(R,0) \equiv 
\sum_{\mu=1}^4\ \langle \Amu^a(\vec{x},0) \Amu^a(\vec{y},0) \rangle 
$
in the MA gauge with the ${\rm U(1)}_3$ Landau gauge fixing.
Here, $\Gmumu^a(R,0)$ depends on the three-dimensional distance 
$R \equiv |{\vec x} - {\vec y}|$ at finite temperature.
In Fig.\ref{caption:Gft},
we show the preliminary results of the numerical simulation 
for $\Gmumu^3(R,0)$ and 
$
G_{\mu\mu}^{+-}(R,0) \equiv 
\sum^4_{\mu=1}\langle A_\mu^{+}(\vec{x},0)A_\mu^{-}(\vec{y},0)\rangle
={1 \over 2} \{G_{\mu\mu}^1(R,0)+G_{\mu\mu}^2(R,0)\}
$ 
in the MA gauge.
Even at finite temperatures,
we find that abelian dominance holds for the 
spatial propagation of gluons in the region of $R \gsim 0.4$ fm.
In Fig.\ref{caption:Gft}, 
$\Gmumu^3(R,0)$ largely changes 
between the confinement and the deconfinement phases.
On the other hand,  $\Gmumu^{+-}(R,0)$ is almost the same 
even in the  deconfinement phase.
The off-diagonal gluon $\Amu^\pm$ in the MA gauge 
seems to propagate within the short-range region as $R \lsim 0.4$ fm 
almost irrespective of the medium of QCD.

\section{Summary and Concluding Remarks}
We have studied the gluon propagator in the MA gauge 
in terms of the interaction range.  
We have found that infrared abelian dominance 
for the gluon propagator in the MA gauge, 
and the interaction range of the off-diagonal gluon  
is restricted within a short distance. 
The off-diagonal gluon behaves as the massive vector boson
with the effective mass $\Meff \simeq 1.2 {\rm ~GeV}$.
Therefore, the off-diagonal gluon $A_\mu^\pm$ can propagate only within 
the  short range as $r \lsim \Meff ^{-1} \simeq 0.2 {\rm ~fm}$, 
and cannot contribute to the infrared QCD physics in the MA gauge. 
Then, abelian dominance holds for the long-distance physics 
with $r \gg \Meff^{-1}$ in QCD in the MA gauge   
and $\Meff^{-1} \simeq 0.2$ fm  can be regarded as 
the critical scale of abelian dominance. 
In this way,  essence of infrared abelian dominance 
in the MA gauge can be physically interpreted with 
the effective off-diagonal gluon mass $\Meff $ induced by the MA gauge fixing,
and the off-diagonal gluon mass generation seems to predict
the general infrared abelian dominance  in QCD in the MA gauge.

\end{document}